\documentclass[a4paper,11pt]{article}
\pdfoutput=1 

\usepackage{jcappub} 

\usepackage[T1]{fontenc} 

\def\calR{\mathcal{R}}
\def\E{\mathbb{E}}
\def\R{\mathbb{R}}

\def\al{\alpha}
\def\be{\beta}
\def\ga{\gamma}
\def\ep{\epsilon}
\def\de{\delta}
\def\vp{\varphi}
\def\ka{\kappa}
\def\si{\sigma}
\def\te{\theta}

\def\Om{\Omega}
\def\La{\Lambda}
\def\Ga{\Gamma}
\def\na{\nabla}
\def\<{\kern-1pt}
\def\arr{\longrightarrow}

\def\Frac[#1/#2]{\frac{#1}{#2}}
\def\({\left(}	\def\){\right)}
\def\[{\left[}	\def\]{\right]}

\title{\boldmath Extended Cosmology in Palatini $f(\calR)$-theories}

\author[a, 1]{P. Pinto\note{Corresponding author.}}
\author[b]{L.Del Vecchio}
\author[b, c]{L.Fatibene}
\author[b]{and M.Ferraris}

\affiliation[a]{Physics Department, Lancaster University, Lancaster LA1 4YB, UK}
\affiliation[b]{Department of Mathematics, University of Torino (Italy)}
\affiliation[c]{INFN - Sezione Torino}

\emailAdd{paolopinto91@gmail.com}
\emailAdd{leonardo.delvecchio@edu.unito.it}
\emailAdd{lorenzo.fatibene@unito.it}
\emailAdd{Marco.ferraris@unito.it}

\abstract{We consider the cosmological models based on Palatini $f(\calR)$-theory for the function $f(\calR)= \al \calR -\Frac[\be/2] \calR^2 -\Frac[\ga/3\calR]$, 
which, when only dust visible matter is considered, is called {\it dune cosmology} in view of the shape of the function $f(\calR(a))$ (being $a$ the scale factor).
We discuss the meaning of solving the model, and interpret it according to the Ehlers-Pirani-Schild framework as defining a Weyl geometry on spacetime.

Accordingly, we extend the definitions of luminosity distance, proper distance, and red-shift to Weyl geometries and fit the values of parameters to SNIa data.
{Since the theoretical prediction is model-dependent, we argue that the fit is affected by an extra choice, namely a model for atomic clocks, which, in principle, produces observable
effects. To the best of our knowledge, these effects have not being considered in the literature before.}

}

\begin{document}
\maketitle
\flushbottom

\section{Introduction}
\label{sec:intro}

We now have significant evidence that standard GR cannot account for observations if we restrict the sources of the gravitational field to the matter we are most familiar with.
This is true now at cosmological scale  (see \cite{DarkSectiorSupernovae}, \cite{CMB}, \cite{WMap}, \cite{PlanckExp}, \cite{FermiExp}), for  almost a century at the scale galaxy cluster dynamics (see \cite{ClusterDM1}, \cite{ClusterDM2}), and, more recently, at the scale of galaxies (see \cite{GalaxyDM1}, \cite{GalaxyDM2}, \cite{Lensing} and \cite{Rev}).

At more or less all scales, maybe with the exception of the Earth and the solar system, one can call for new sources we are not able to see directly yet (which are collectively named {\it dark sources})
or, alternatively, for a modification of {the} dynamics of {the} gravitational field (i.e.~the prescription of how ordinary visible matter generates a gravitational field). 
It is clear that {\it a priori} the two approaches are equally possible (and somehow equivalent) until we have a direct evidence of dark sources  other than gravitational ones,
that, currently, we do not have.

{In terms} of modifications of dynamics, we have a (too) rich list of proposals: conformal gravity, MOND, metric, Palatini, and metric-affine $f(R)$-theories, torsion, Lovelock, just to
quote {a few}.  Each one has proven well in some specific situation, but none have yet proven to be a solution in general.
{Hence, one needs a generic and robust framework to test models with a  solid control on observational protocols, which often are not uniquely determined by the action 
principle, but they are extra choices one does to connect to observations.}

{For} modifications of sources, in cosmology, the {\it$\La$CDM}(-concordance) model has been accepted as a good description of current observations.
Currently, observations require the cosmic pie to be  $\Om_\La\simeq 0.70$ {\it dark energy} (in the form of a cosmological constant $\La$), $\Om_c\simeq 0.25$ of some {\it dark matter} which we do not see though it shares with ordinary matter the same equation of state (EoS) and of which we have no local, fundamental, direct evidence other than its gravitational effects,  $\Om_b\simeq 0.05$ of ordinary baryonic matter which accounts for the {\it visible matter} we see in form of galaxies and gas, as well as traces $\Om_r\simeq 10^{-4}$  of {\it radiation} (and relativistic matter) which have very little effect today even though, due to a different scaling property, they grow important in an earlier universe.
The {\it$\La$CDM}(-concordance) model also has no spatial curvature $k=0$.

The {\it$\La$CDM}-model is considered the best description of current observations, a kind of standard which any other proposed model must reproduce.
One also often {\it believes} that the dark matter added from cosmological evidence is the same dark matter one needs in galaxy and cluster dynamics,
an identification for which, however, a detailed quantitative parallel is missing and it is currently theoretically out of reach.
{To state} the obvious, extrapolating matter equations of state (EoS), which are already an approximation, from a cosmological scale down to a galactic scale is brave, {not even} mentioning the lack of knowledge behind the fact we do not know an elementary counterpart for dark matter.

Of course, the same bravery is needed in trying to extend the same dynamics from human scale up to cosmology, not to mention down to the quantum world.
Furthermore, often most observations rely on a gravitational model for interpretation and that such an interpretation is often quite fragile with respect to modification of assumptions about geometry  of spacetime.
We have to admit that, beside raw data, we do not know much for sure and currently any study needs extra care.

Among modifications of gravitational dynamics, we shall here consider a particular class, namely the {\it extended theories of gravitation} (see \cite{OlmoNostro}, \cite{ETG}, \cite{ETC}).
In extended theories of gravitation, one has a Weyl geometry on spacetime, i.e.~$(M, g, \tilde \Ga)$, instead of the usual Lorentzian metric structure.
The (torsionless) connection $\tilde \Ga$ is {\it a priori} independent of the metric $g$. While the connection describes the free fall of test particles (and light rays) 
in the gravitational field, the metric $g$ is chosen to simplify the gravity-matter coupling and (consequently) to account for atomic clocks and, in turn, our protocols for measuring distances.

The dynamics are chosen so that the metric and the connection, {\it a priori} independent, turn out to be {\it a posteriori}, i.e.~as a consequence of field equations, 
{\it EPS-compatible} (see \cite{EPS}), which means there exists a $1$-form $A= A_\ep dx^\ep$ such that
\begin{equation}
\tilde \Ga^\al_{\be \mu} = \{g\}^\al_{\be \mu}  -{\frac{1}{2}} \(g^{\al\ep}g_{\be\mu} - 2\de^\al_{(\be} \de^\ep_{\mu)}\) A_\ep
\end{equation}

This framework is theoretically inspired and motivated by a work on foundations of gravitational physics by Ehlers-Pirani-Schild (EPS); see \cite{EPS}.
The application to extended theories is described in \cite{EPSNostro}, \cite{ETG}, \cite{ETC}, \cite{Roshan}.
 
In an extended gravitational theory, one has modifications of dynamics which can equivalently be seen as effective sources. Extended theories also contain 
standard GR, with or without a cosmological constant, as a special (quite degenerate) case.

A class of dynamics which are automatically extended theories of gravitation are Palatini $f(\calR)$-theories (see below and \cite{reviewPalatinif(R)}), in which, field equations not only imply EPS-compatibility, 
but they also imply that $A$ is closed or, equivalently, that $\tilde \Ga$ is metric, i.e.~$\tilde \Ga=\{\tilde g\}$ for a metric $\tilde g$ which is conformal to $g$.
Being $g$ and $\tilde g$ conformal, they define the same pointwise causal structure, the same light cones, the same light-like geodesics.
However, they define different timelike geodesics so that it is important to declare that $\tilde g$, rather than  $g$, defines the free fall of test particles.
Also, being conformal, the $g$-length of a $\tilde \Ga$-parallelly transported vector is not preserved, though at least it depends on the point only, not on the curve along which is parallelly transported.

In this simpler case of two conformal metrics, the extra (kinematical) freedom one has with respect to standard GR is encoded in the {\it conformal factor} $\vp$, a scalar (real, positive) field
such that $\tilde g= \vp \cdot g$.
We have to stress, however, that, in view of the specific form of the action functional in Palatini $f(\calR)$-theories,
all these objects are not dynamically independent. For example, once we know the field $\tilde g$, then the conformal factor $\vp$ as well as the metric $g$
are uniquely determined as function of $\tilde g$ and its derivatives (up to order two). 
They are not extra physical degrees of freedom, for example in the sense that one cannot ``excite'' one without exciting the others.
The dynamics of $\vp$ and $g$ (or $\tilde g$) are uniquely determined once the dynamics of $\tilde g$ (or $g$) is given, as {it} usually happens to Lagrangian multipliers.
The metric $g$ shares with Lagrangian multipliers the fact that it enters the gravitational Lagrangian with no derivatives (so, in a sense, its field equations are algebraic).

Consequently, finding a solution in a Palatini $f(\calR)$-theory actually means determining all $\tilde g$, the conformal factor $\vp$, as well as the original metric $g$.
Of course, one could recast the action functional in terms of purely $g$ or $\tilde g$, though at the price of making the matter--gravity coupling more complicated
and messing up with the interpretation of the theory about which we made a clear choice: times and distances are measured with $g$, free fall with $\tilde g$.

{ Traditionally, doing all with the metric $g$ is referred as the {\it Jordan frame}, while using $\tilde g$ is called the {\it Einstein frame}.
In the model we study in this paper, we do not use either the Jordan or the Einstein frame. We argue instead that one should not expect either of the metrics to be used for everything, as it happens in standard GR, and that is the essence of Weyl geometries. 
It is clear, for example by EPS kinematic analysis, that free fall and the causal structure are structures coming from different physical phenomena (free fall is associated to test particles, causality to light rays) and one has no reason to assume {\it a priori} a constraint between them.
The choice of $\tilde g$ for geodesics and $g$ for causal and metric structures is precisely what makes this model different from the other analyses in which one frame is chosen to describe both structures.
We shall eventually argue that this feature turns out to be in principle observable and important when one is going to compare the model with the solar system classical tests.
}

\medskip
The main aim of this paper is to discuss the application of a specific model of a Palatini $f(\calR)$-theory  to cosmology.
We discuss how the interpretation of the gravitational physics is extended from a Lorentzian metric geometry, to the more general Weyl conformal geometry, 
{Furthermore, we investigate} how {the} observations are interpreted in this more general setting  (see \cite{Perlick}), which contains the standard GR case as a special case, as we said. 
If one does not like the model on a physical stance, one can regard this paper as a proposal for setting a rigorous standard for interpretation of observations in cosmology
as well as an example of how a model should be discarded from an observational stance.
As a matter of fact, Palatini $f(\calR)$-theories are naturally candidates to be at least a setting for understanding tests of GR in a wider context, something which was originally done with Brans-Dicke theories for historical reasons (see \cite{Weinberg}) while we are suggesting it should be done in extended gravity.

\section{Notation}

We hereafter consider a cosmological model with a dynamics based on a Palatini framework, 
i.e.~fundamental fields $(g_{\mu\nu}, \tilde \Ga^\al_{\be\mu})$  and an action  functional
\begin{equation}
A_D(g, \tilde \Ga, \psi)= \int_D \(\frac{\sqrt{g}}{2\ka} f(\calR) + L_m(g, \psi)\) d\si
\qquad\qquad
\ka:=  \frac{8\pi G}{c^3}
\label{Action}\end{equation}
where $\psi$ denotes matter fields and $\sqrt{g}d\si$ is the volume element induced by the metric $g$.
The quantity $\sqrt{g}$ is the usual square root of the absolute value of the determinant of the metric tensor.
The function  $f(\calR)$ will be here chosen as
\begin{equation}
f(\calR)= \al \calR -\Frac[\be/2] \calR^2 -\Frac[\ga/3\calR]
\label{fR}\end{equation}
where we set $\calR :=g^{\mu\nu} \tilde R_{\mu\nu}$ and $ \tilde R_{\mu\nu}$ is the Ricci tensor of the (torsionless) connection $\tilde \Ga$.

{ The choice of function (\ref{fR}) has been not very thoughtful. 
If one sets $\gamma=0$ the master equation becomes too easy to be inverted, but one has no late acceleration. 
Moreover, here we want to set up a framework able to deal with more generic $f(\calR)$, rather then choosing one for which we can easily do the computations.
Under this viewpoint, the function (\ref{fR}) shows a number of pathologies which are good to learn to cope with, at least in a classical regime.
On the other hand, we do not want to add too many terms, keeping the degeneracy minimal. The more parameters one adds the easier the fitting becomes, the more experiments one should use to actually remove degeneracy and to really constrain the parameters.

If we had chosen $\ga=0$, we would obtain a sort of Starobinsky model (see \cite{Staro}), though in Palatini formalism. That is simpler to analyse (the master equation is globally and analytically invertible). However, in this simplified model, one will have no negative pressure in effective EoS, no late time acceleration, the conformal factor will be asymptotically constant. Still, also in this model, one has a bouncing rather than an initial singularity (in the Jordan frame).

} 

Similar models have been considered in the purely metric formulation, see \cite{Nojiri:2017ncd} \cite{Od1}, \cite{Od2}, \cite{Od3}.
{It has been argued that these models are not viable as physical models, mainly for stability issues and classical tests; see \cite{Amendola1}, \cite{Amendola2}, \cite{Dolgov}, \cite{Chiba}.
On this basis, further models have been proposed in the metric formalism to address these shortcomings; see \cite{Straro}, \cite{Appleby}, \cite{Hu}.
Even if, in view of the non-equivalence between the metric and the Palatini formulation, the same critiques do not apply directly to our model,
we believe it is useful to briefly review them in a purely metric context to pinpoint what they exactly disprove.

Most of the models proposed deal with cosmology only, where they used Jordan frame. 
All the critiques deal with stability and solar system tests (or, equivalently, Newtonian limit) in the Jordan frame as well.
They show in many cases the Jordan frame is not viable in the solar system.
However, in cosmology, one uses only light and comoving test particles which, as a consequence of cosmological principle, are shared by the Jordan and Einstein frames.
As a consequence, in cosmology (at least until perturbations are considered) one never really uses the metric $\tilde g$. 
If it is certainly true that solar system tests do not allow $g$-geodesics to describe test particles, however, it is also true that assuming that test particles are described by $\tilde g$
does not change anything in the cosmological model and it gives a {\it different} solar system model (which, by the way, can pass the tests quite easily, in view of the universality theorem; see \cite{Universality}).

Accordingly, we completely agree that (metric) $f(R)$-theories in the Jordan frame (at least the ones considered in the references above) are not compatible with solar system tests.
We just need to mention that when doing cosmology, one is not really compelled to declare the frame and there are mixed models, in which both $g$ and $\tilde g$ are used to do different things, which have not been discussed explicitly.
It is our opinion they should be discussed and possibly disproven as well.
Let us stress that, in $f(\calR)$-theory there is no dynamical equivalence between purely metric and Palatini formalisms, even at the level of a simple counting of
physical degrees of freedom. Accordingly, disproving metric models leaves Palatini models unchallenged.

}

Of course, one could argue that, because of the form of the action functional, this model is certainly non-renormalisable and that the Minkowski metric is not even a solution.
Of course, this is true also for standard GR with a cosmological term. It may be that the model is not well suited for quantum gravity.
However, on one hand, we do not know what quantum gravity will eventually be precisely or whether it will require renormalisable theories or it will rather be non-perturbative in nature. 
On the other hand, we are here discussing a {\it classical} model, which unfortunately has nothing to do (observationally speaking) with the quantum regime.
And there are many ways a classical model with a singular Lagrangian can well behave, especially if the standard for well-behaviour is what happens in standard GR,
in which  singularities are already bound to appear generically.
Kepler motion in a plane is another example of a singular Lagrangian which is accepted to describe a well-behaving (mechanical) system, in which the conservation of angular momentum prevents, most of the times, the system to get to the singularity.

As far as the fact that Minkowski is not even a solution, the identification of Minkowski, and only Minkowski spacetime, as the vacuum state of gravitational field is already quite dubious at a fundamental level. A theory involving a metric field has no canonical vacuum just because metrics (or vielbein) do not carry a linear (or affine) structure.
Metric theories are different from all other fundamental field theories. Already standard GR is a peculiar field theory in which one should learn to live without many of the structures used in field theories in Special Relativity (SR). For example, in GR one has no linear structure for configurations, generically no Killing vectors, no fixed background.

In what follows, we shall assume that the connection $\tilde \Ga$ is responsible for free fall. Particles will follow geodesic trajectories of $\tilde \Ga$.
The metric $g$ is related to distances on spacetime and its causal structures, e.g.~the light cones. 
For example, a freely falling atomic clock will follow a timelike geodesics worldline with respect to $\tilde g$, though the parameterisation is chosen to be
proper with respect to $g$. 
Of course, the difference is expected to be tiny, though we have to keep in mind that we wish then to discuss objects going around for $3\cdot 10^{17}s$, with plenty of time to grow the tiny difference until it may become appreciated.
Extrapolation at scales by many order of magnitudes requires good definitions and possibly no mathematical approximations.

If experience still eventually points in favour of standard GR dynamics, we shall have obtained it without relying on unnecessary theoretical assumptions, but based on
experience and a better understanding of which assumptions we rely on.

In {the} literature, there are not many studies for Palatini $f(R)$-theories; see \cite{Borowiec}, \cite{Olmo2011} and references quoted therein.
{ See also \cite{Szydlowski1}, \cite{Szydlowski2}, \cite{Szydlowski3} for polynomial models}.
This is often argued to be due to a number of {problems that  Palatini $f(R)$-theories} are supposed to have which have been however {refuted}; see \cite{no-go},  \cite{EqBD}. 
We shall not discuss here these issues since they are discussed in \cite{Olmo}, \cite{Mana}, \cite{Wojnar}, \cite{ETG}, \cite{MathEquivalence}.

Field equations for the action (\ref{Action}) are obtained by varying with respect to $\de g^{\mu\nu}$, $\de \tilde \Ga^\al_{\be\nu}$, and $\de \psi^i$:
\begin{equation}
\begin{cases}
 f'(\calR) \tilde R_{\mu\nu} -\frac{1}{2} f(\calR)g_{\mu\nu} =\ka T_{\mu\nu} \cr
\tilde\na_\al (\sqrt{g} f'(\calR) g^{\be\mu})=0\cr
\E_i =0\cr
\end{cases}
\end{equation}
In general, the second is solved by defining a conformal factor $\vp= (f'(\calR))^{\frac{m-2}{2}}$, $m$ being the dimension of spacetime, a conformal metric $\tilde g_{\mu\nu}= \vp g_{\mu\nu}$
and by showing that $\tilde \Ga=\{\tilde g\}$ is thence the general solution of the second field equation (which, written in terms of $\tilde g$ and $\tilde \Ga$,
is actually algebraic, in fact linear, in $\tilde \Ga$).

{ The third equation $\E_i =0$ is obtained as a variation of the action with respect to the matter fields $\psi$. It describes how matter fields evolve in the gravitational field.
Usually, in cosmology, one does not give a precise Lagrangian description of the matter dynamics, which is described (under additional assumptions) by EoS, thanks to which Friedmann equations become well-posed. Accordingly, we shall neglect the specific form for it.}

By tracing the first equation by means of $g^{\mu\nu}$, one obtains the so-called {\it master equation}
\begin{equation}
f'(\calR) \calR -\frac{m}{2} f(\calR) =\ka T
\end{equation}
where we set $T:= g^{\mu\nu}T_{\mu\nu}$.
This is also an algebraic equation in $\calR$ and $T$ which generically can be (at least locally) solved for $\calR= \calR(T)$, so that the curvature $\calR$ along solutions can be expressed as a (model dependent but) fixed function of the matter content $T$.
 
At this point, the first field equation can be recast as the Einstein equation for the metric $\tilde g$ (or, equivalently, for the conformal metric $g$)
\begin{equation}
\tilde R_{\mu\nu} -\frac{1}{2} \tilde R \tilde g_{\mu\nu} = \ka \tilde T_{\mu\nu} 
\qquad\iff\qquad
 R_{\mu\nu} -\frac{1}{2}  R  g_{\mu\nu} = \ka \hat T_{\mu\nu} 
\label{EE}\end{equation}
In both cases, the energy--momentum stress tensors ($\tilde T_{\mu\nu}$ or $\hat T_{\mu\nu}$) need to be modified by sending to the right hand side all spurious contributions from matter (or curvature). Let us stress that also $\hat T_{\mu\nu}$ differs from the original $T_{\mu\nu}$
which instead is the usual variation of the matter Lagrangian with respect to the metric $\de g^{\mu\nu}$. We shall not use $\hat T_{\mu\nu}$, while let us mention that
\begin{equation}
\tilde T_{\mu\nu} := \Frac[1/f'(\calR)] \( T_{\mu\nu}  - \Frac[ f'(\calR) \calR - f(\calR) /2\ka]  g_{\mu\nu}\)
\end{equation}
This is where effective {\it dark sources} come from in Palatini $f(\calR)$-theories.
Whatever {\it visible matter} is, it is described by $T_{\mu\nu}$, then $\tilde T_{\mu\nu}$ directly gets extra contributions from the modified dynamics, i.e.~from the function $f(\calR)$
which, hopefully, by choosing it accordingly, can be used to model dark matter and energy as effective sources.
This is not the only effect in extended theories. Also the odd definition of atomic clocks (which are free falling with respect to $\tilde g$ but proper with respect to $g$) 
produces extra accelerations in particles. These accelerations are universal, i.e.~they are easily confused with an extra gravitational field acting on all test particles equally which, when reviewed in a standard GR setting, calls for other sources. Hereafter, we shall investigate the combination of these two types of effects in cosmology.

{ It is precisely because we chose $\tilde g$ to describe test particles that we are not working in the Jordan frame, 
and because we chose $g$ to describe clocks that we are not working in the Einstein frame, either.
That is true, even though, when we restrict to cosmology, one could argue that we are working in the Jordan frame since the comoving structure is shared by those two frames.
However, the Einstein frame pops out again if we go to discuss gravity in the solar system, where choosing the frame to describe test particles leads to different models.}

\section{Extended cosmologies}

Let us consider a four dimensional spacetimes with a Weyl geometry $(M, g, \{\tilde g\})$.
If we want to build a cosmological model based on the extended theories described above, we need to impose  {the} cosmological principle.
Of course, with two metrics, one should at least stop and think which metric should obey the cosmological principle.
The good news is that (since the master equation holds) it does not matter: $g$ is spatially homogeneous and isotropic iff $\tilde g$ is.
The only difference is that, {if $g$ is in FLRW} form in coordinate $(t, r, \te,\phi)$, with a scale factor $a$, then $\tilde g$ is in FLRW form in coordinate $(\tilde t, r, \te,\phi)$, with a scale factor $\tilde a= \sqrt{\vp}\> a$. If the conformal factor is a function only of time, the new time is defined by $d\tilde t = \sqrt{\vp} dt$.

Thus one has a Friedmann equation both for $a$ and $\tilde a$
\begin{equation}
\dot a^2 = \Phi(a)
\qquad\qquad
\dot {\tilde a}^2 = \tilde \Phi(\tilde a)
\label{FE}\end{equation}
which are, of course, defined to be equivalent.
The specific form of the function $\Phi(a)$ and $\tilde \Phi(\tilde a)$ are obtained by expanding the Einstein equations (\ref{EE}), which are, in fact, equivalent.

As a consequence of the cosmological principle, the energy-momentum tensor $T_{\mu\nu}$ is in the form of a perfect fluid energy-momentum tensor, namely
\begin{equation}
T_{\mu\nu} = c^{-1} \( (\rho c^2+p) u_\mu u_\nu + p g_{\mu\nu}\)
\end{equation}
for some time-like, future directed, $g$-unit, comoving vector $u^\mu$. 
Also $\hat T_{\mu\nu}$ can be recast in the same form (for different $\hat \rho$ and $\hat p$), as well as $\tilde T_{\mu\nu}$ is a perfect fluid energy-momentum 
tensor using $\tilde g$ and a suitable $\tilde g$-unit vector $\tilde u$ as well as different effective pressure and density $\tilde p$ and $\tilde \rho$.

To establish an equivalence between Einstein equations and Friedmann equation, we need conservation of the relevant energy-momentum tensors. Luckily enough, once again, if $\na_\mu T^{\mu\nu}=0$ (as it is, since it is variation of a covariant matter Lagrangian) then $\tilde \na_\mu \tilde T^{\mu\nu}=0$ and $\na_\mu \hat T^{\mu\nu}=0$ as well. Here, $\na_\mu$ denotes the covariant derivative with respect to $g$, $\tilde \na_\mu$ the covariant derivative with respect to $\tilde g$.

Finally, we need to state {the} EoS for matter. This is where the game becomes odd: imposing the EoS for visible matter, i.e.~for $p$ and $\rho$ appearing in $T_{\mu\nu}$,
the EoS for $\tilde \rho$ and $\tilde p$ in $\tilde T_{\mu\nu}$ are uniquely determined (as well as the EoS for $\hat \rho$ and $\hat p$ in $\hat T_{\mu\nu}$).
However, ``simplicity'' is not preserved. Even if we assume visible matter to be simply dust (i.e.~we select $p=0$ as EoS) then the EoS for effective matter is determined, though the effective EoS is very exotic and non-linear. Even if we regard it as a mixture of simple polytropic fluids, the decomposition is not canonical and, in any event, it contains different
polytropic fluids. Again, since we wish to discuss a model at cosmological scale, it is necessary to avoid mathematical approximations in EoS and learn to live with what we have, even when it is complicated to compute.

As {the} usual in cosmology, one normalises the scale factor to be unit today, i.e.~$a(t_0)=1$. Since the conformal factor is defined up to a constant factor which does not {affect} {Christoffel} symbols $\{\tilde g\}$, one can also normalise the conformal factor to be (positive and) $\vp(t_0)=1$, so that the conformal transformation preserves
the (positivity and) normalisation of the scale factor and one has also $\tilde a(\tilde t_0)=1$.

Regardless which equation of  (\ref{FE}) we decide to solve, once we have $t(a)$, then the know the conformal factor as a function of $a$ from the master equation, so we also have $\tilde a(a)$.
Then we also know $\vp(a)$ and hence $\tilde t(a)$, $\rho(a)$, $p(a)$, $\tilde \rho(a)$, $\tilde p(a)$ and so on.
We get everything as a function of $a$ so all quantities are known in parametric form as function of every other (and no need to invert functions, other than the master equation).

For an energy-momentum tensor in the form of a perfect fluid, we have $T= c^{-1}(3p-\rho c^2)$ and, for simplicity, we set {the} EoS for visible dust $p=0$.
That, in view of energy-momentum tensor conservation, is equivalent to set $\rho(a) := \rho_0 a^{-3}$.

If we fix the function (\ref{fR}), the master equation reads as
\begin{equation}
 \al \calR-\be \calR^2 +{\frac{\ga}{3}} \calR^{-1}  - 2 \( \al \calR -\frac{\be}{2} \calR^2 -\frac{\ga}{3} \calR^{-1} \)
 = - \al \calR  +  \ga \calR^{-1} = \Frac[\ka/c] (3p- \rho c^2)
\end{equation}
which can be solved in two branches (corresponding to the sign of $\calR$) as
\begin{equation}
{}^\pm \calR(a) =  \Frac[ \ka \(\rho c^2 - 3 p\)\pm \sqrt{ \ka^2 \(\rho c^2-3p\)^2+4c^2 \al  \ga} /2c\al ] 
\end{equation} 

If we consider a mixture of dust ($p_d=0$) and radiation ($p_r=\frac{1}{3} \rho_r c^2$)
\begin{equation}
\rho c^2-3p = \rho_d c^2-3p_d + \rho_r c^2 -3p_r
 = \rho_d c^2 + \rho_r c^2 -\rho_r c^2
  = \rho_d  c^2
\end{equation}
Thus we have $\rho_d = \rho_0 a^{-3}$ and, consequently
\begin{equation}
{}^\pm \calR(a) =  \Frac[ \ka c^2 \rho_d \pm \sqrt{ \ka^2c^4 \rho_d^2+4c^2 \al  \ga} /2c\al ] 
=  \Frac[ \ka c \rho^d_0  \pm \sqrt{ \ka^2 c^2 \(\rho^d_0\)^2 +4 \al  \ga a^{6}} /2\al a^{3}] 
\label{R(a)}
\end{equation}

In case one wants different types of visible matter, though, the extra pressure would need to be taken into account.
We shall show {the} result for {the} values:
\begin{equation}
{\al \simeq 0.095
\qquad\qquad
\be=0.25\> m^2
\qquad\qquad
\ga \simeq  2.463\cdot 10^{-104} \>m^{-4}  }
\end{equation}

{
When we discuss fits in Section 5,
we shall argue that if we analyse this model to fit SNIa there is a lot of degeneracy.
The fit is not really able to constrain the parameters $\al$ and $\be$, though if their values are provided by some other test (e.g.~solar system tests) then
supernovae will fix $\ga$.
Still, the analysis of the model is interesting because having enough tests to remove the degeneracy. 
At this point, we are introducing a small $\be$ and we are not here concerned with the small value of $\al$.
See discussion in the conclusions.
}

The conformal factor is chosen to be proportional to $f'(\calR)$
which is everywhere positive if we use ${}^- \calR(a)$, while ${}^+ \calR(a)$ changes sign at about {$\rho_1:=1.925\cdot 10^{24} \> kg\> m^{-3}$}.
Thus, for the conformal factor to be positive, we need to define it in three branches

\begin{itemize}

\item{--} the branch {A}, with $\calR>0$ and and $\rho\in(\rho_1, +\infty)$  (thus $a\in (0, a_1)$), where the conformal factor is defined as
$\vp_A := -\vp_0 f'(\>{}^+\<\calR)$;

\item{--} the branch B,  with $\calR>0$ and $\rho\in(0,\rho_1)$  (thus $a\in (a_1,+\infty)$), where the conformal factor is defined as
$\vp_B := \vp_0 f'(\>{}^+\<\calR)$;

\item{--} the branch {C}, with $\calR<0$ and $\rho\in(0,+\infty)$ (thus $a\in (0,+\infty)$), where the conformal factor is defined as
$\vp_C := \vp_0 f'(\>{}^-\<\calR)$;
\end{itemize}
\noindent
where $\vp_0$ is a constant to be chosen so that today $\vp(t_0)=1$.

Branch A corresponds to very high densities, so it happened early in the universe. We assume then to currently be on branch $B$ at $a=a_0=1$.
So we choose $\vp_0 := \( f'(\>{}^+\<\calR(a_0=1)) \)^{-1}$.

The effective (mass) density and pressure are 
\begin{equation}
\tilde \rho =\Frac[  4 \ga\calR -3 \be \calR^4 + 12\ka c \rho\calR^2 / 4\ka c \(3 \al \calR^2-3 \be \calR^3+\ga\) \vp]
\qquad\qquad
\tilde p= -\Frac[ 4c \ga\calR -3c \be \calR^4  -12\ka  p\calR^2 / 4\ka  \(3 \al \calR^2-3 \be \calR^3+\ga\) \vp]
\end{equation}
where $\rho$ and $p$ are the total mass density and pressure of visible matter.  

If we have only visible dust, $\rho=\rho_d$ and $p=0$.
If we have visible dust and radiation, then $\rho= \rho_d+\rho_r$, $p=p_r=\frac{1}{3}\rho_r c^2$.

We also clearly see that effective sources in general cannot be dust, as well as cannot be polytropic.

By using the correct expression for $\vp$ and $\calR$ on each branch, we can compute {the} Friedmann equation
\begin{equation}
\dot {\tilde a}{}^2 = \frac{\ka c^3}{3} \tilde \rho(\tilde a)\> {\tilde a}^2 -k c^2  =: \tilde \Phi (\tilde a)
\end{equation}

In view of the transformation between the two frames induced by the conformal factor, we have
\begin{equation}
\dot a ^2= \Phi(a) :=   \vp(a)  \({\Frac[d\tilde a/d a]}\)^{-2} \tilde \Phi (\tilde \rho(a)) 
=   \vp(a)  \({\Frac[d\tilde a/d a]}\)^{-2}  \(\Frac[{\ka {c^3}}/3]   \tilde \rho(a) \tilde a^2(a)  -kc^2\)
\end{equation}

{ Hence, the  Friedmann equation evaluated today reads as 
\begin{equation}
\omega^2 H_0^2 = \(\Frac[{\ka {c^3}}/3]   \tilde \rho(\rho_0)   -kc^2\)
 \qquad\qquad  
\(\omega :=\Frac[d\tilde a/ d a](1)\)
\end{equation}
Since the Hubble parameter today  $H_0$ is measured, then we can obtain the spatial curvature as a function of the visible matter density $\rho_0$, i.e. 
\begin{equation}
 k(\rho_0) = c^{-2} \(\Frac[{\ka {c^3}}/3]   \tilde \rho(\rho_0) -\omega^2 H_0^2 \)
 \label{krho}
\end{equation}
Let us remark that once the function $f(\calR)$ is fixed, then we know the function $\tilde \rho(\rho)$ and the constant $\omega$.
}

Accordingly, on each branch we can compute the function $\Phi(a)$, exactly, depending on the parameters $(\al, \be, \ga, \rho_0)$ of the theory. 
In Figure \ref{fig:1}.b~in the Appendix we draw the graph of the function $\Phi(a)$ for branches A and B.

{ For any given value of the spatial curvature $k$ one can compute the corresponding value of the density which produces it. 
As usual the {\it critical density} is the density which produces a spatially flat spacetime $k=0$.
Thus we have a 3-parameter family of extended models, all with the observed value of the Hubble parameter today.}

One can explicit the time as a function of $a$ solving the integral
\begin{equation}
t(a) = \int_1^a \Frac[da / \sqrt{\Phi(a)}]
\label{Solution}
\end{equation}
{ Then the parametric curve $\ga: a \mapsto (t(a), a)$ represents the graph of the function $a(t)$.
Let us notice that, in this way, one can study the function analytically, at the price of a finite number of numerical integration, even when the integral cannot be performed analytically.}

For realistic parameters, we obtain the evolution of the scale factor (dark-solid), compared with {\it$\La$CDM} (light-solid) and standard GR (light-dashes).

\begin{figure}[htbp] 
   \centering
   \includegraphics[height=5cm]{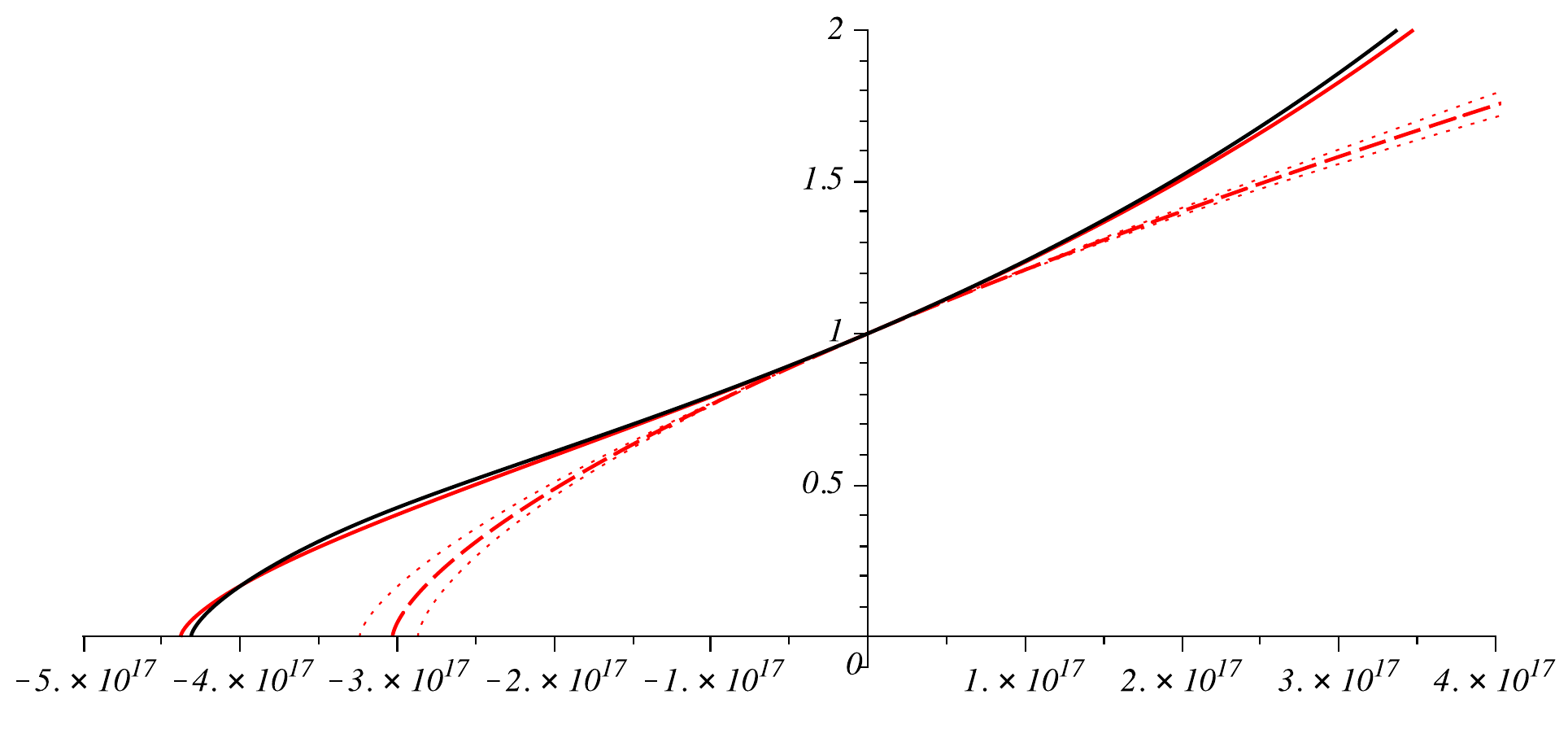} 
   \caption{\small\it 
   Evolution of $a(t)$ in standard GR for critical density (light dash), 
   standard GR for $\pm 30\%$  of the critical density (light dotted),
   {\it$\La$CDM} (light solid), 
   and dune cosmology for the critical density (dark solid).
   Times in seconds ($x$-axis), $a$ is adimensional ($y$-axis).}
   \label{fig:H}
\end{figure}

The 3 models are almost identical near today $(t=0)$ while they differ in the past and in the future. In particular, the extended $f(\calR)$ model exhibits, for the critical density, a slightly younger age for the universe (with a Universe age of about {${13.68}By$}).

Once we solved the model, then we can obtain all other quantities as a function of $a$.
Qualitative graphs are collected in Appendix A.

\section{Distances in extended theories}

To fit data from supernovae ({\it SNIa}), we need a precise definition of the luminosity distance $d_L$, the proper distance $\de$, and the red-shift $z$
of a source within our model.
In particular, we need to extend the standard discussion which is based on a Lorentzian geometry to an (integrable) Weyl geometry $(M, g, \{\tilde g\})$.

 In cosmology, one defines spatial distances as the geometric distance on the surface $t=t_0$, without relying on synchronisation of clocks, which, of course, would be impractical in astrophysics since easily it would  take millions (if not billions) of years for a signal to bounce back and forth from another galaxy.
Any comoving observer, at its time $t=t_0$, defines a surface $t=t_0$ and chooses a geodesics on the surface (with respect to the metric induced on the surface by $g$).
The {\it proper distance} $\de(t_0)$ is then the $g$-length of such a (space-like) geodesics.

Let us remark that the geodesic on the surface is in general not a geodesic on spacetime, as a geodesic on a sphere $S^2$ in $\R^3$ is not a straight line. 

The surface $t=t_0$, which is here defined using the time coordinate, {can also} be defined in terms of the Killing algebra of isometries prescribed by the cosmological principle in an intrinsic fashion.

The proper distance is a geometric well-defined distance, though it is difficult to define a protocol to measure it directly. We use it as a benchmark to refer the luminosity distance $d_L$ and the red-shift $z$ of a source.

In our model, light rays move along geodesics of $\tilde g$ which are light-like with respect to $g$. However, since $g$ and $\tilde g$ are conformal, they share the same
light-like geodesics. So we can consider light-like geodesics using only the metric $g$. Accordingly, the red-shift is given by 
\begin{equation}
z(a) = \Frac[a_o- a/a] 
\end{equation}
where $a_e=a$ is the scale factor of $g$ at emission, $a_o$ at observation. If we make observations today, of course we have $a_o=1$.
The scale factor $\tilde a$ of the metric $\tilde g$ does not play a role in observations until we assume that atomic clocks are proper for $g$ and not for $\tilde g$,
though they free fall along $\tilde g$-geodesics.

This can also be shown in detail, directly by repeating the standard argument in a Weyl geometry; see \cite{book2}.

Similarly, the area of the sphere $S^2$ for $t=t_0$ and $r=r_\ast$, measured by the metric $g$ is exactly $A= 4\pi a^2(t_0) r_\ast^2$.
Accordingly, the luminosity distance of a comoving source at $r=r_\ast$ observed at $t=t_0$ is 
\begin{equation}
d_L= (1+z) a_o r_\ast
\end{equation}
which, if $k=0$, is exactly $d_L= (1+z) \de(t_o, r_\ast)$.
If one wishes not to assume $k=0$, depending on the sign of $k$, $r_\ast$ is anyway a known function of the proper distance.

There is no mathematical approximation we made here, in particular, 
we are not restricting ourselves to near sources by using a linear approximation of the Hubble law.

Now we know $z(a)$ as a function of the emission scale factor. If we observe today sources at different distances, we have
their proper distance $\de(t_o, r_\ast)$ as a function of the emission scale factor $a$
\begin{equation}
\de(a) :=  c\int_a^{a_0} \Frac[ da / a\sqrt{\Phi(a)}]
\end{equation} 
Given that we have $d_L(a)$ and $z(a)$, we have a parametric representation of the function $d_L(z)$.

That function depends on the dynamics of the model through the Weierstrass function, namely  $\Phi(a; \al,\be,\ga,\rho_0)$, so that it brings information about the model, and the function $f(\calR)$ in particular.

For this reason, we can fit the parameters to obtain a best fit representation of the observed curve.

\section{Fitting the Ia type supernovae}

Since we know that the relationship between the magnitude of a far standard electromagnetic source (e.g. Ia type supernovae) and the observed redshift is determined by the dynamics of the scale factor $a(t)$, it is possible to use statistical inference methods to evaluate the agreement between the theoretical prediction within our model and the experimental observations.
The complexity of the modelling of both the theory and observations requires correspondingly refined statistical and data analysis skills. In fact, the measurements of the magnitude and the redshift of the SNIa must account for a strong uncertainty signal in the background (see \cite{SNLS}), usually described by two nuisance unknown parameters $a$ and $b$.
For this reason, a Bayesian inference approach is often used in this case (but more generally in all the cosmological measurements). 

We have decided to perform two different sets of fit for our model using the data concerning the measurement of the Ia type supernovae magnitude as a function of the observed redshift. We considered the Supernovae Legacy Survey (SNLS) project   catalogue composed by 115 SNIa (see \cite{SNLS}) and the whole Union2.1 catalogue  (see \cite{Union2}) composed by 580 SNIa. We have a clear and unambiguous relation between these physical observables and the mathematical objects in the theory, which makes these datasets particularly well suited for model-theoretic parameters fit.
{ The SNLS is a smaller dataset than Union2.1 though its data are more homogeneous, so we use it as a check for consistency. }

The theoretical value of the magnitude $m$ of a source as a function of its redshift $z$ has been calculated considering the flow of power carried by the momentum-energy tensor associated to a high frequency electromagnetic wave propagating in a homogeneous and isotropic spacetime. One can notice that the explicit form of $m(z)$ is strongly related to the dynamics of the universe so it depends on the initial conditions (e.g.~the baryonic and radiation energy density today, the Hubble parameter today) as well as the vacuum Lagrangian parameters that determine the evolution of metric tensor $g$.  

To fit the SNIa data, we relied on the software MULTINEST (see \cite{Mu1}, \cite{Mu2}, \cite{Mu3}), an efficient and robust Bayesian inference tool developed to calculate the evidence and obtaining posterior samples from distributions with (an unknown number of) multiple modes and pronounced (curving) degeneracies between parameters. 

The power of the software lies in the algorithm that naturally identifies the individual modes of a distribution, allowing for the evaluation of the local evidence and parameter constraints associated with each mode separately.
The fit was performed asking MULTINEST to find the free parameters of the theory $(\alpha,\beta,\gamma,\rho_0; a, b)$ that minimizes the $\chi^2$ defined as following:
\begin{equation}
\chi^2= \sum_{i=1}^N \Frac[[m_{Bi}-m(z^\star_i)]^2 / \sigma(m_B)^2+\sigma^2_{int}]
\end{equation}
where:
\begin{itemize}
\item{-} $ z^\star_i$ is the measured redshift;
\item{-} $m_{Bi}= m_i- a(s-1)+bc$ is the cleaned real magnitude in which: $m_i$ is the measured magnitude, $s$ is the stretch factor, $c$ is the colour factor, $a$ and $b$ are free nuisance parameters that are fixed by the fit;
\item{-} $m(z^\star_i)$ is the theoretical prediction of the magnitude at given redshift;

\item{-} $\sigma(m_B)^2=\sigma(m)^2+\sigma(s)^2+\sigma(c)^2$ is the error of the observations. In particular, we have that $\sigma(m)^2$ is the error related to the magnitude, $\sigma(s)^2$ and $\sigma(c)^2$ are respectively the errors of the stretch factor and color factor;
\item{-} $\sigma^2_{int}=0.13104$ the error related to the intrinsic dispersion of the real SNIa from the ideal standard candle.
\item{-} $N$ is the number of observed supernovae belonging to the dataset, 115 for the SNLS dataset, 580 for the Union2.1 dataset.
\end{itemize}

MULTINEST is also able to provide us with the value of the $\chi^2$ evaluated on the best fit parameters, the posteriors samples and the live points produced by the algorithm. This is very useful both for checking the right convergence to the minimum and to estimate the posterior probability distribution of our parameters. The posteriors analysis as well as the confidence region has been performed using the software GetDist; see \cite{GetDist}.

Furthermore, considering the minimum value of the $\chi^2$, we can compare different theoretical models and determine the accuracy of the theoretical predictions.

\medskip
Different cases have been studied: at first, we have considered the case in which all the parameters {$(\al, \be, \ga, \rho_0, a, b)$ are fitted against the Union2.1 dataset.
That fit shows that the model is strongly degenerate, that the $\beta$ parameter does not influence the scale factor at the scales sampled by the dataset as well as that $\al$ is very poorly localized.}

Then we analysed cases in which some parameters are set to a value determined by different considered scenarios.
{That shows that if we have some value for parameters from tests other than SNIa, the model is still able to fit supernovae, for example to determine the $\ga$ parameter (as well as $a$ and $b$). This shows how SNIa are unable to fix all parameters of the model, as one can reasonably expect. 
The same behaviour in some sense is noticed in standard GR, where the parameters $(\al, \rho_0)$ are fixed at different scales, $\al$ being fixed by the solar system tests, and 
$\rho_0$ by supernovae.

Here we expect something similar, only with more parameters. We expect different tests at different scales (solar system, light elements formation, \dots)
to remove the degeneracy and fix best fit value for the parameters. Then, and only then, the model can be tested to predict new phenomena (e.g.~BAO, lensing, \dots).
}

\begin{figure}[htbp] 
   \centering
   \includegraphics[height=9.5cm]{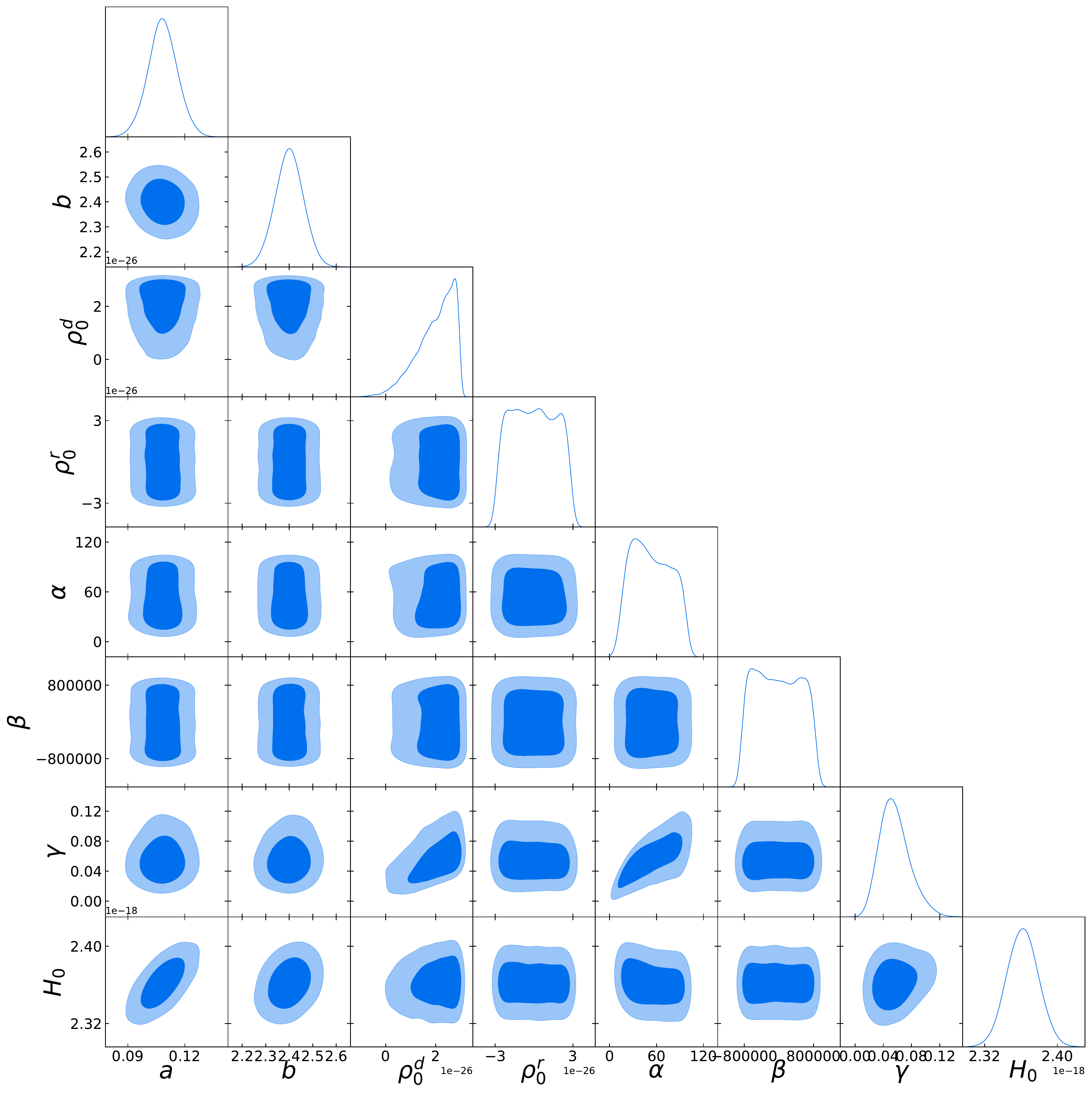} 
   \caption{\small\it
   Triangular plot of the posterior distribution generated from the chains of MULTINEST.\protect\\
   The coloured shapes represent the regions of the parameter space with a confidence level of $2\sigma$ and $1\sigma$. \protect\\
    We fit all parameters $(\rho_0^d, \rho_0^r, \al, \be, \ga, H_0, a, b)$. We can see many parameters (e.g. $\al$ and $\be$) are not very well constrained by the fit.}
   \label{fig:A1}
\end{figure}

\medskip
{The first fit we tried is using the Union2.1 dataset, fitting all parameters $(\al, \be, \ga, \rho^d_0, \rho^r_0, H_0, a, b)$.
At each step in the simulation, we compute the spatial curvature $k$ using equation (\ref{krho}).

The best fit parameters are
\begin{equation}
\begin{cases}
\al= 11.996 ^{+87.962}_{-9.599} \\
\be= -0.097^{+ 9.284}_{-9.480} \cdot 10^{16}  \>  m^{2}\> Mpc^{2}\> km^{-2}=  -0.092^{+8.840}_{-9.025} \cdot 10^{55} \> m^{2}\\
\ga= 2.178^{+ 8.661}_{-1.753} \cdot 10^{-24} \> km^4\>  m^{-4}\>Mpc^{-4}=0.024 ^{+0.095}_{-0.019}\cdot 10^{-100} m^{-4} \\
\end{cases}
\end{equation}
The matter content is determined as
\begin{equation}
 \rho^d_0=1.375_{-1.160}^{+1.621} \cdot 10^{-26} \> kg\> m^{-3}, \quad
  \rho^r_0=2.526 _{-5.523}^{+0.470}\cdot 10^{-26} \> kg\> m^{-3}
\end{equation}
the Hubble parameter as
\begin{equation}
 H_0=73.106_{-1.376}^{+1.215} \> km\> s^{-1}\> Mpc^{-1}=2.369_{-0.045}^{+0.039}\cdot 10^{-18} \> s^{-1}, 
\end{equation}
and the nuisance parameters as
\begin{equation}
 a=0.108^{+0.019}_{-0.020},\quad
 b= 2.398^{+0.144}_{-0.133}
\end{equation}
The spatial curvature is computed as
\begin{equation}
 k= -1.958^{+2.030}_{-2.439} \cdot 10^{-53} m^{-2} \> 
\end{equation}
The fit has $\chi^2_R= 0.370$ which could indicate an overestimation of the errors.
See Figure \ref{fig:A1} for the triangular plot.

}

\begin{figure}[htbp] 
   \centering
\includegraphics[height=7.5cm]{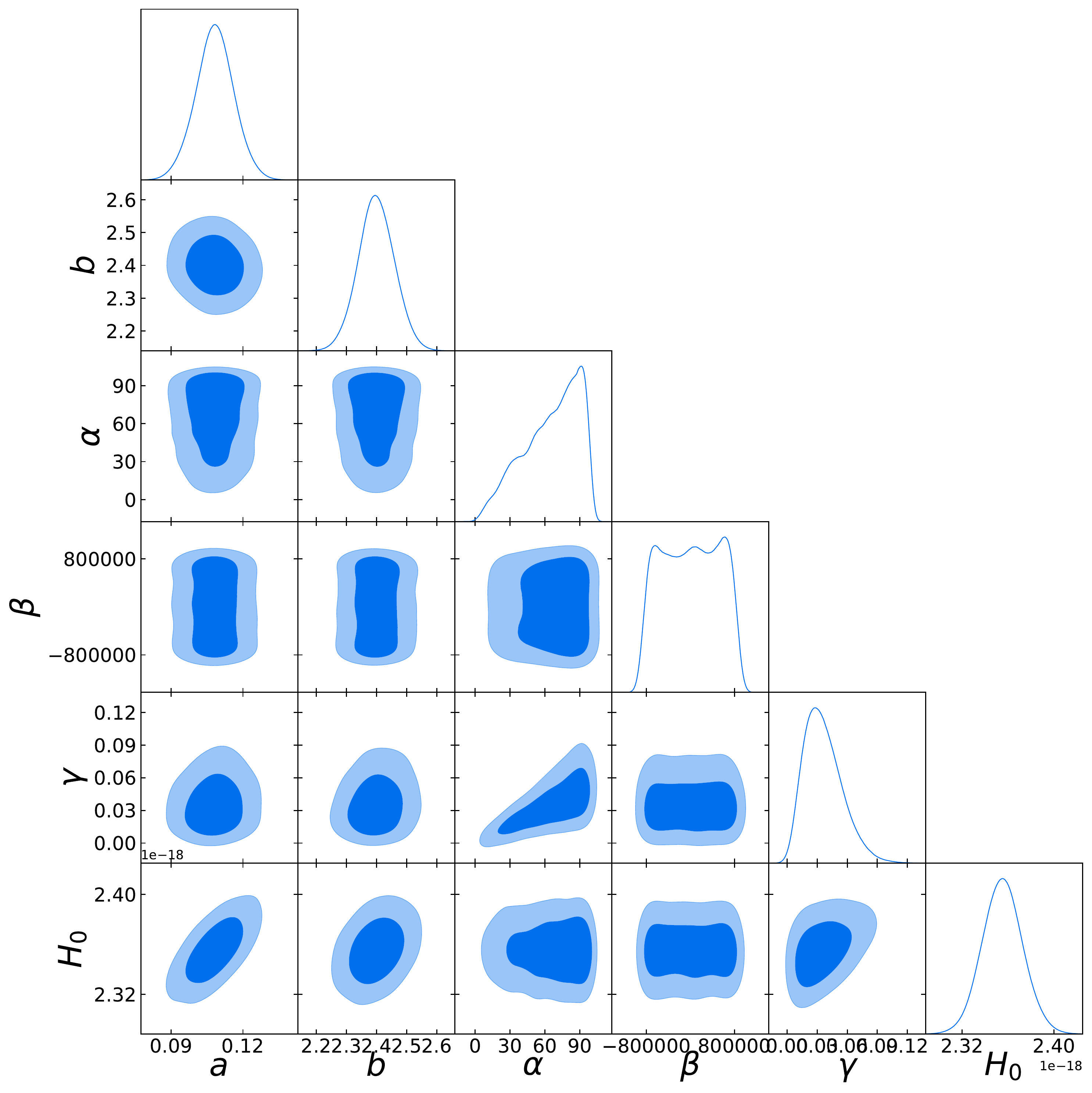}
   \caption{\small\it
   Triangular plot of the posterior distribution generated from the chains of MULTINEST.\protect\\
   The coloured shapes represent the regions of the parameter space with a confidence level of $2\sigma$ and $1\sigma$. \protect\\
   We fix $\rho_0^d$ and $\rho_0^r$ (to the density they have in {\it$\La$CDM} for the sake of discussion) 
   and fit $(\al, \be, \ga, H_0, a, b)$. The values of $\al$ and $\be$ are still poorly constrained.}
   \label{fig:A2}
\end{figure}

\medskip
{ The goodness of the fit is not particularly relevant here, since the best fit parameters confidence intervals indicate that, as one could expect, the system is quite degenerate and some of the parameters, $\al,\be,\ga$ in the first place, are quite poorly constrained. 
 This fit is also compatible with negative densities of radiation which are of course unphysical. Let us notice that, first of all, one can interpret it as saying that SNIa are not able to constrain radiation, i.e.~that, as far as only SNIa are concerned, the radiation density could be zero. As for removing degeneracy, one will need further tests to fix radiation contribution. Secondly, we are interested here in showing that the extended model we are discussing will be potentially able to fit data with the same visible sources and no dark sources at a fundamental level. This is a long process, it will take into consideration many different tests as explained in Conclusions. In the following fits concerning SNIa, we shall fix a radiation density similar to the one predicted in the {\it$\La$CDM} model, since SN themselves are not able to constrain the value of $\rho_0^r$.

The second fit we try is using Union2.1 dataset, fixing $\rho_0^d= 0.418 \cdot 10^{-27}kg \>m^{-3}$,  $\rho_0^r= 0.001 \cdot 10^{-27} kg \>m^{-3}$, 
and fitting all other parameters $(\al, \be, \ga, H_0, a, b)$.
At each step in the simulation, we compute the spatial curvature $k$ using equation (\ref{krho}).

The best fit parameters are
\begin{equation}
\begin{cases}
\al= 0.369 ^{+99.627}_{-0.171} \\
\be= -7.825^{+ 17.210}_{-1.561} \cdot 10^{16} \>  m^{2}\> Mpc^{2}\> km^{-2}=  -7.450^{+16.388}_{-1.486} \cdot 10^{55} \> m^{2}\\
\ga= 6.540^{+ 837.97}_{-1.740} \cdot 10^{-26} \> km^4\>  m^{-4}\>Mpc^{-4}=7.214 ^{+924.329}_{-1.926}\cdot 10^{-104} m^{-4} \\
\end{cases}
\end{equation}
The Hubble parameter is determined as
\begin{equation}
 H_0=73.106_{-1.376}^{+1.215}\> km\> s^{-1}\> Mpc^{-1}= 2.369_{- 0.052}^{+0.024}\cdot 10^{-18} \> s^{-1}, 
\end{equation}
and the nuisance parameters as
\begin{equation}
 a=0.107^{+0.015}_{-0.015},\quad
 b= 2.408^{+0.112}_{-0.134}
\end{equation}
The spatial curvature is computed as
\begin{equation}
 k= -2.033^{+2.986}_{-1.000} \cdot 10^{-53} m^{-2} \> 
\end{equation}
The fit has $\chi^2_R= 0.368$.
See Figure \ref{fig:A2} for the triangular plot.

}

\begin{figure}[htbp] 
   \centering
   \includegraphics[height=8.5cm]{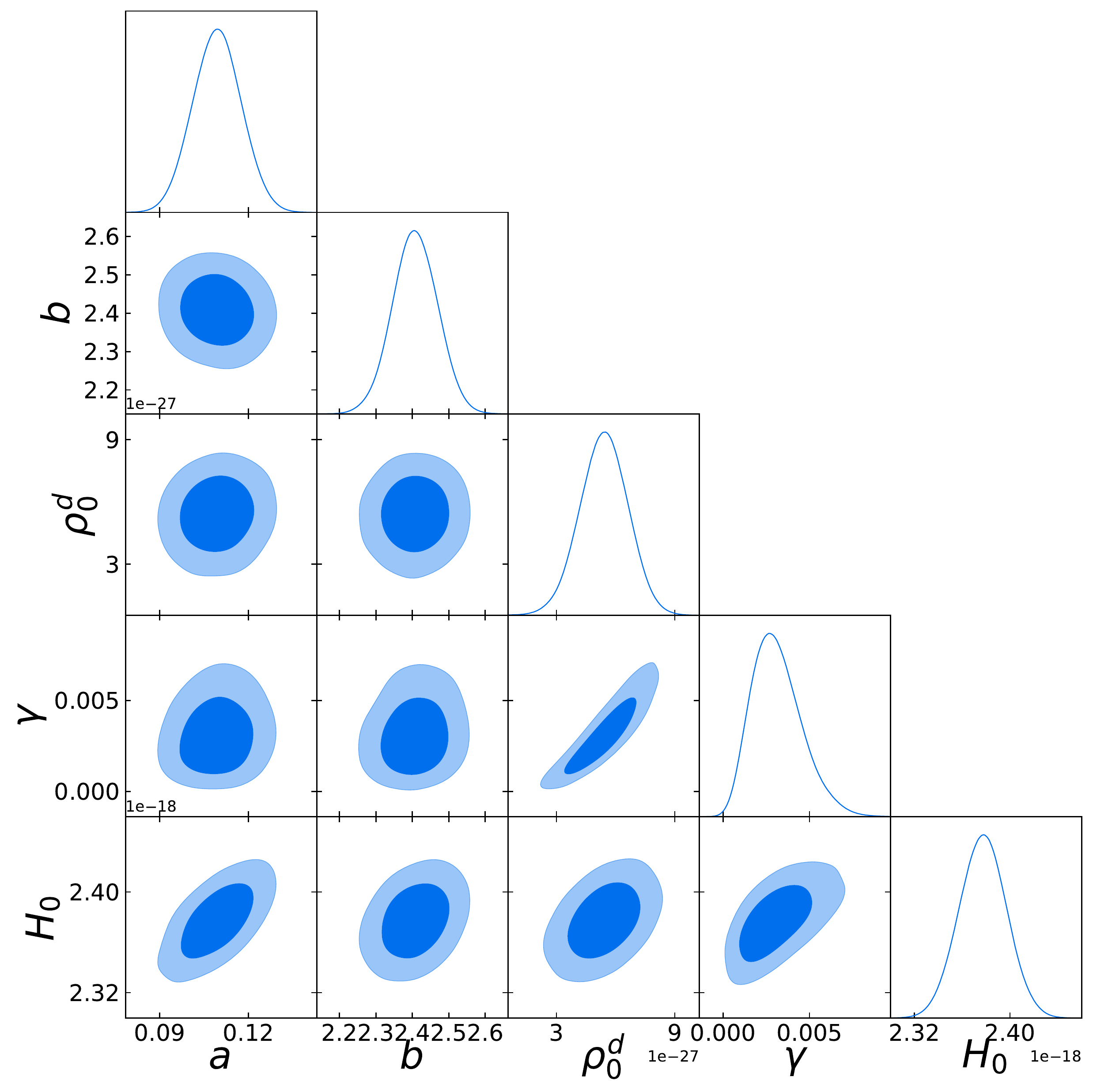} 
   \caption{\small\it
   Triangular plot of the posterior distribution generated from the chains of MULTINEST.\protect\\
   The coloured shapes represent the regions of the parameter space with a confidence level of $2\sigma$ and $1\sigma$. \protect\\
   We now fix  $\alpha=1$, $\be=0$, $\rho_0^r$, fitting $(\rho_0^d, \ga, H_0, a, b)$. 
   This time the constraint of values by the fitting is more convincing.}
   \label{fig:A3}
\end{figure}

\medskip
{ Also providing the densities for visible matter, the parameters of the model are still poorly constrained.
Thus, as a third fit we fix $\al=1$ and $\be=0$, as well as  $\rho_0^r= 0.001 \cdot 10^{-27} kg \> m^{-3}$.
Then, still using the Union2.1 dataset, we fit all the other parameters $(\rho_0^d, \ga, H_0, a, b)$, while still computing $k$ at each step.

The best fit parameter is
\begin{equation}
\ga= 2.417^{+ 3.706}_{-2.346} \cdot 10^{-25} \> km^4\>  m^{-4}\>Mpc^{-4} =2.667 ^{+ 4.089}_{-2.588}\cdot 10^{-103} m^{-4} \\
\end{equation}
The dust density and the Hubble parameter is determined as
\begin{equation}
\begin{cases}
 \rho^d_0=5.212_{-3.689}^{+2.647} \cdot 10^{-27} \> kg\> m^{-3} \\
 H_0=73.259_{-1.292}^{+1.442}  \> km\> s^{-1}\> Mpc^{-1} =2.374_{-0.041}^{+0.046}\cdot 10^{-18} \> s^{-1}
\end{cases}
\end{equation}
and the nuisance parameters as
\begin{equation}
 a=0.108^{+0.018}_{-0.018},\quad
 b= 2.408^{+0.134}_{-0.137}
\end{equation}
The spatial curvature is computed as
\begin{equation}
 k= -0.423^{+3.500}_{-4.466} \cdot 10^{-53} m^{-2} \> 
\end{equation}
The fit has $\chi^2_R= 0.367$.
See Figure \ref{fig:A3} for the triangular plot.

 }

\medskip

\begin{figure}[htbp] 
   \centering
\includegraphics[height=8.5cm]{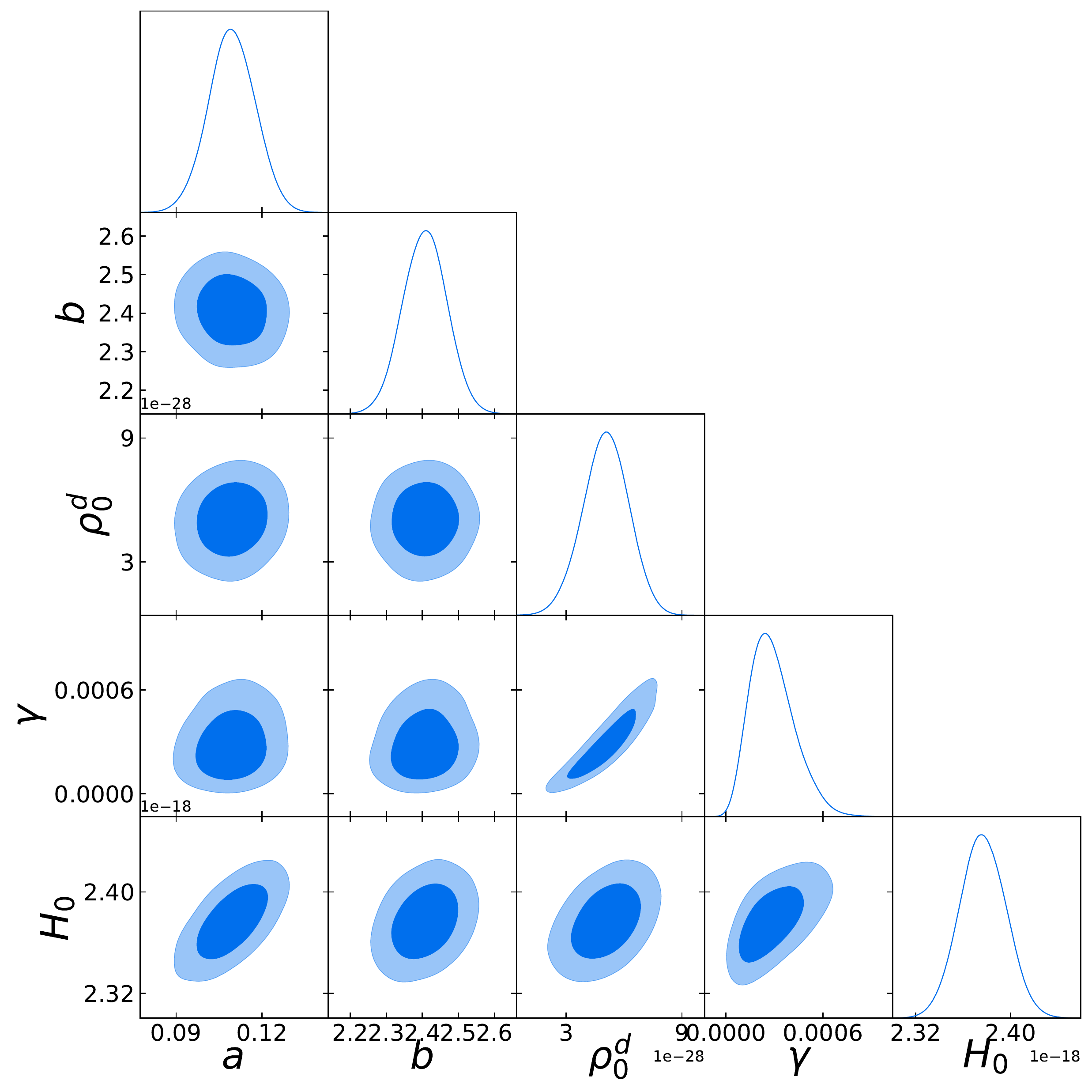}
   \caption{\small\it
   Triangular plot of the posterior distribution generated from the chains of MULTINEST.\protect\\
   The coloured shapes represent the regions of the parameter space with a confidence level of $2\sigma$ and $1\sigma$. \protect\\
    We now fix  $\alpha=0.095$, $\be=0$, $\rho_0^r$, fitting $(\rho_0^d, \ga, H_0, a, b)$. 
   This time the dust density is compatible with {\it$\La$CDM}.
}
   \label{fig:A4}
\end{figure}

{ This time we see that the fit is better convincing. The only issue with respect to {\it$\La$CDM} is that the dust density is (about one order of magnitude) higher than
in {\it$\La$CDM}. At this stage it is not clear whether this is a prediction of the model or it is simply due to the value imposed on the other parameters.
For, let us consider  a fourth fit, where we fix $\al=0.095$ and $\be=0$, as well as  $\rho_0^r= 0.001 \cdot 10^{-27} kg \> m^{-3}$.
Then, always using Union2.1 dataset, we fit all other parameters $(\rho_0^d, \ga, H_0, a, b)$, still computing $k$ at each step.

The best fit parameter is
\begin{equation}
\ga= 2.223^{+3.479}_{-2.034} \cdot 10^{-26} \> km^4\>  m^{-4}\>Mpc^{-4}=2.463 ^{+3.838}_{-2.244}\cdot 10^{-104} m^{-4} \\
\end{equation}
The dust density and the Hubble parameter is determined as
\begin{equation}
\begin{cases}
 \rho^d_0=0.489_{-0.304}^{-0.257} \cdot 10^{-27} \> kg\> m^{-3} \\
 H_0=73.242_{-1.364}^{+1.328} \> km\> s^{-1}\> Mpc^{-1}=2.373_{-0.044}^{+0.043}\cdot 10^{-18} \> s^{-1}
\end{cases}
\end{equation}
and the nuisance parameters as
\begin{equation}
 a=0.109^{+0.018}_{-0.017},\quad
 b= 2.404^{+0.141}_{-0.134}
\end{equation}
The spatial curvature is computed as
\begin{equation}
 k= -0.505^{+3.497}_{-4.459} \cdot 10^{-53} m^{-2} \> 
\end{equation}
The fit has $\chi^2_R= 0.367$.
See Figure \ref{fig:A4} for the triangular plot.

In these cases the space curvature is about of the same order of magnitude of experimental constraints ($k\sim 10^{-54}m^{-2}$ from Planck), 
just with a bigger uncertainty, as one can expect since here we used only SNIa data.
In any event, it is compatible with spatial flatness.

\medskip
We see that in this fourth fit the obtain more or less the same dust density as in {\it$\La$CDM}, thus confirming that one can change $\al$ and $\rho_0^d$
accordingly, so that the fit adjusts the value of $\ga$ but still it is a good fit. That explicitly highlights the degeneracy we originally guessed.

We also checked what happens in the third and fourth fits by restricting to the SNLS dataset.
The SNLS is smaller, so we expect higher errors, though it is more homogeneous, so we might prevent systematic errors in calibration. 
For this reason, it is interesting to check that one obtains about the same results as with Union2.1
dataset.

We obtained best fit values which are completely compatible with the ones of the Union 2.1 catalogue, both with reduced chi-squared values of {$\chi^2_R=1.390$}.

 }

The last fits are reasonably good (maybe they show a overestimation of the errors). They show us the way to test the model. 
One needs more tests, for example solar system tests and light elements formation, to fix $\al$ and $\be$.
Then, with that extra information, supernovae seem to be able to determine the parameter $\ga$ (as well as the matter content $\rho_0^d$, the Hubble parameter $H_0$,
and the nuisances $a$ and $b$).

{ Let us stress that the value of $\ga$ is not compatible with $\ga=0$ showing that this model is not a simple deformation of Starobinsky model $f(\calR)=\al\calR+ \be \calR^2$. 
The model we are considering has late time acceleration, but supernovae do not constrain $\be$ which in some fit is even considered $\be=0$.}

For the best fit values we found in the fourth fit, we showed the evolution of the scale factor $a(t)$, see Figure \ref{fig:H}.
In Appendix A, we provide qualitative graphs for all relevant functions in the model.

\section{Conclusions and perspectives}

{We can conclude that we have many degrees of freedom and degeneracy in the parameters, too many and too much to determine all  parameters just by fitting the SNIa. 
However, if we impose some theoretical constraints, for example by tests at different scales, we are always able to determine at least the value of $\gamma$, as well as $H_0$, $a$ and $b$, in different scenarios. More tests are needed. Tests are model dependent and each new test must be done within a model framework.
It is dangerous and reckless to guess the results on the basis of an intuition which has been developed in standard GR.

Only after one has enough tests to remove the degeneracy and fix the model, then new phenomena will allow us to discuss the predictivity of the theory.
Currently, we believe that solar system test can fix $\al$, while light elements production could fix $\be$.
Only at that point could one really discuss the physical content of the theory, for example by discussing BAO, CMB, structures formation,
gravitational lensing, galaxies and clusters dynamics (see \cite{Guzzo}),
for each of which one still needs to develop a complete treatment within the framework of $f(\calR)$-Palatini theories.

The specific model we consider here produces late time acceleration, it is compatible with SNIa observations (as many other models are, including {\it$\La$CDM}),
it predicts a cosmological dynamics which is quite closed to that of {\it$\La$CDM} (see Figure \ref{fig:H}), yet it is observationally different in principle.
That is simply not enough to propose or dismiss it as a sound physical model.

Even if we did not believe that the model $f(\calR) = \al \calR -\Frac[\be/2] \calR^2 -\Frac[\ga/3]\calR^{-1}$ is a sound  physical proposal, it certainly shows that Palatini $f(\calR)$-theories are potentially able to model SN observations and cosmology without adding dark sources at a fundamental level.
Also, we showed that  Palatini $f(\calR)$-theories are observationally different from standard GR and {\it$\La$CDM}.

This approach is also interesting because it poses a new standard for tests, with respect to the traditional comparison with Brans-Dicke models, which is now much better founded on first principles, e.g.~in view of EPS.
Let us remark that if standard GR, in order to fit observations, needs to account for dark sources (at all scales)  at a fundamental level, 
then alternative, modified or extended theories of gravitation need to account (thoroughly and in full detail) for how observations arise in their framework.
One cannot simply say that  a model fits observations without dark sources, and not explaining in details where the extra accelerations (which are usually interpreted
as the effect of the gravitational field produced by dark matter) come from.

Here we considered a specific Palatini $f(\calR)$-theory, we assumed an interpretation for the fields $g$, $\tilde g$, and matter fields, which supports EPS framework
and provides us with a solid bridge between the model and observational protocols.  These assumptions are not proven, they are part of the model specification and 
can be falsified or corroborated by experiments. That is the correct way of proceeding.

From our analysis, it seems that either $\al$ is much smaller than expected or dust density is much higher than usually supposed.
Within this model,if a smaller value of $\al$ is not supported by observations, then either one has more dust around or the model is contradicted by current cosmological data.
One could also argue that one needs the dark matter contribution to the dynamics ($\Omega_b+\Om_c\simeq 0.30$) for non-cosmological reasons (galaxies and cluster dynamics) which are quite well established.
However, one should say that to discuss galaxy models in a Palatini $f(\calR)$-theories, one should also consider the effect of conformal factor, 
that in these models is expected to depend on $r$, not (only) on cosmological $t$. 
Is that enough to fit observations without adding dark matter?

}

We are also planning to devote future investigations to describe, the evolution of the Hubble parameter $H(z)$ as a function of distance (being the distance parameterised by $z$, $a$, or $t$) within this model.
That will provide a test when the new data for Hubble drift will be available and, at least, it provides new evidence that Palatini $f(\calR)$-theories can be, in principle,  falsified by the observations.

We are currently working to split the effects in extended theories, into the component due to effective sources from the effects due to the atomic clocks being proper with respect to a metric which is not the one describing free fall of test particles.
Our interpretation is well based in EPS framework, though being able to split the effects, will enable us to test this assumption by experiments.

{  Also a full analysis of the dynamical system describing the cosmological sector of this theory would be interesting, as done in \cite{Szydlowski3} for polynomial models.
Here, we rather analysed the system around a specific set of parameters, while a full analysis would highlight whether one can have finite time singularities for other values of the parameters. For now, we know no finite time singularity arises for our best fit parameters.
}

Of course, one would need also perturbation theories, structure formation, lensing, interactions with particle models (e.g.~baryogenesis), the evolutions of perturbations of CMB
and many other aspects. The point is that proceeding in that direction is the only way to falsify models on a certain basis.

{And that is the only thing we can do, until we are able to discuss model properties without fixing the function $f(\calR)$, something which currently is completely out of reach.}

\section*{Appendix $A$:  Qualitative graphs}

The model is solved as soon as we get $t(a)$, see (\ref{Solution}). All other quantities can be computed as functions of $a$ so that they all can be plotted with respect to all the others in parametric form.
For realistic parameters, graphs are difficult to be analyzed since the interesting features occur at many order of magnitude away one from the others.
So here we collect the graphs for the parameters values
\begin{equation}
\ka=c=1
\qquad 
\al=1
\qquad 
\be=\frac{1}{16}
\qquad 
\ga=\frac{1}{128}
\qquad 
\rho_0=\frac{3}{4}
\end{equation}
These graphs show qualitatively the form of model relations with realistic parameters.

\begin{figure}[htbp] 
   \centering
{   \includegraphics[width=6cm]{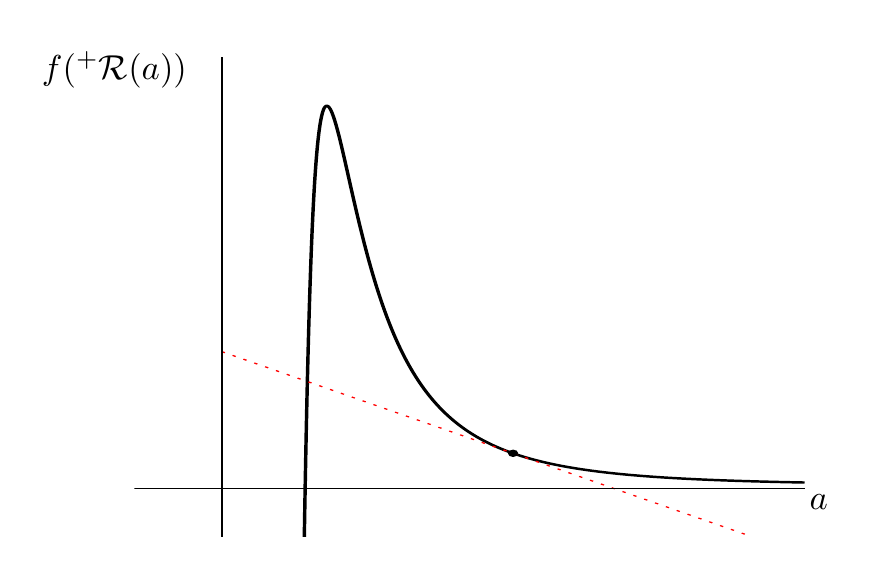}  }
{ \includegraphics[width=4cm]{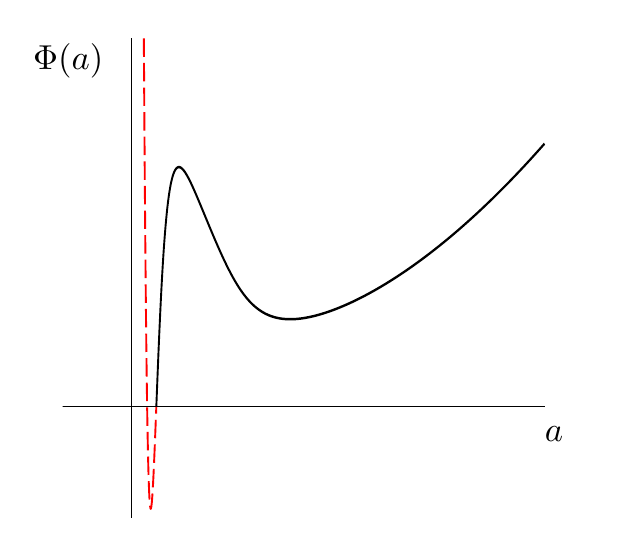}  }
{   \includegraphics[width=4cm]{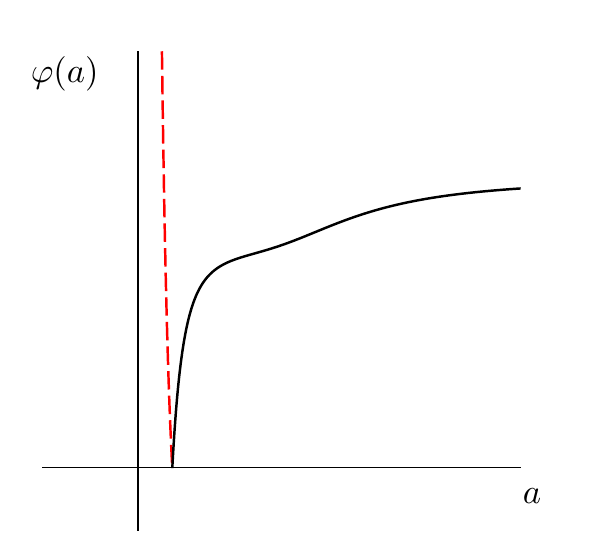}  }
   \caption{\small\it The graphs show in order:\protect\\
   a) the scalar density $f({}^+\calR(a))$ as function of the scale factor $a$\protect\\
   b) the Weierstrass function $\Phi(a)$ as function of the scale factor $a$\protect\\
   c) the conformal factor $\vp(a)$ as function of the scale factor $a$
    }
   \label{fig:1}
\end{figure}

In view of the master equation and EoS for visible matter, equation (\ref{R(a)}) expresses the curvature $\calR$ as a function of the scale factor $a$.
The function $f(\calR)$ is given as a function of $a$ as $f({}^+\calR(a))= 0.7388 -2.124 (a-1)+O((a-1)^2)$ in Figure \ref{fig:1}.a.
That corresponds, near today, to approximately standard GR with a positive cosmological constant.
Going back in time the action climbs the {\it dune} and falls down again unlike in {\it$\La$CDM}. 
Also the limit for $a\arr +\infty$ is different from the standard case in which $\calR \propto a^{-3}$.
For this reason, we call this model {\it dune cosmology}.

{The Weierstrass function $\Phi(a)$, which determines the Friedmann equation for the scale factor $a$, for these parameters is shown in Figure \ref{fig:1}.b.
One can see that there is a bounded allowed region $(0, a_1]$ (corresponding to a universe which readily recollapses)
as well as an unbounded one $[a_2, +\infty)$ in which we are now.
The scale factor $a=a_2$ acts as a reflection point (i.e.~a bouncing point). The second derivative of the Weierstrass function acts as a driving force.
Accordingly, a positive second derivative indicates an accelerating expansion while around the maximum we have a deceleration phase.}

\begin{figure}[htbp] 
   \centering
{   \includegraphics[height=4cm]{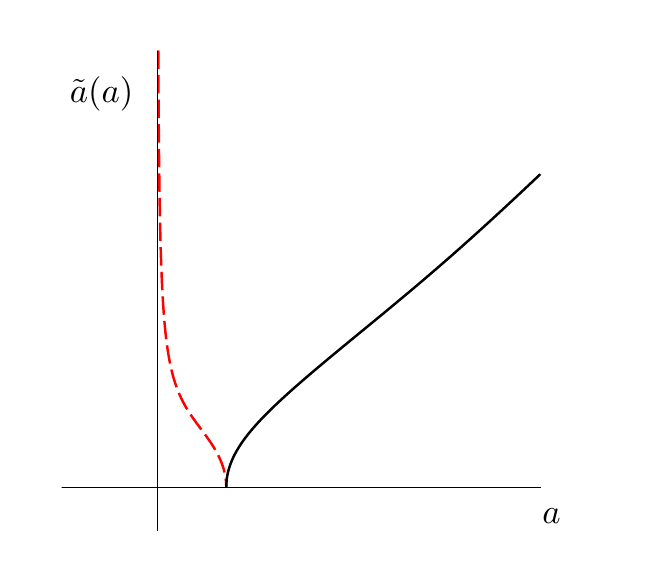}  }
 \hfill
{   \includegraphics[height=4cm]{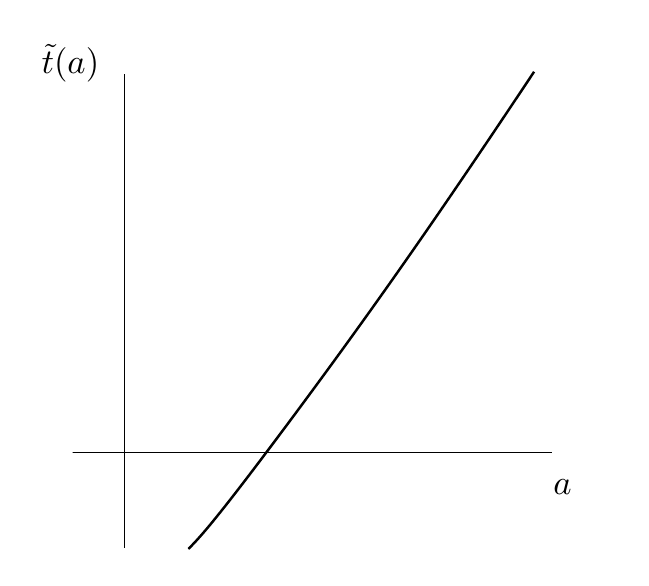}  }
 \hfill
 {   \includegraphics[height=4cm]{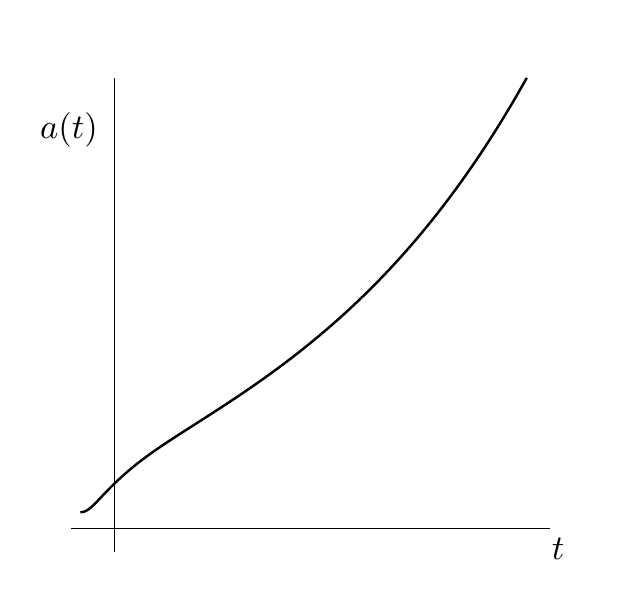}  }
   \caption{\small\it The graphs show in order:\protect\\
   a) the scale factor $\tilde a(a)$ of the metric $\tilde g$ as function of the scale factor $a$\protect\\
   b) the coordinate time $\tilde t(a)$ as function of the scale factor $a$\protect\\
   c) the evolution of the scale factor $a(t)$ as a function of time $t$
   }
   \label{fig:2}
\end{figure}

The conformal factor $\vp(a)$ is presented in Figure \ref{fig:1}.c, $\tilde a(a)= \sqrt{\vp(a)}\> a$ in Figure \ref{fig:2}.a, and finally $\tilde t(a)$ in Figure \ref{fig:2}.b,
{which is obtained by integrating the equation 
\begin{equation}
\frac{d\tilde t}{da} = \sqrt{\vp}\> \frac{dt}{da} = \> \sqrt{\frac{\vp(a)}{\Phi(a)}}
\end{equation}}

Then we integrate $t(a)$ (in Figure \ref{fig:2}.c) which in fact exhibits an initial slowing down phase, followed by an accelerated expansion.

\begin{figure}[htbp] 
   \centering
\hfill{   \includegraphics[height=4cm]{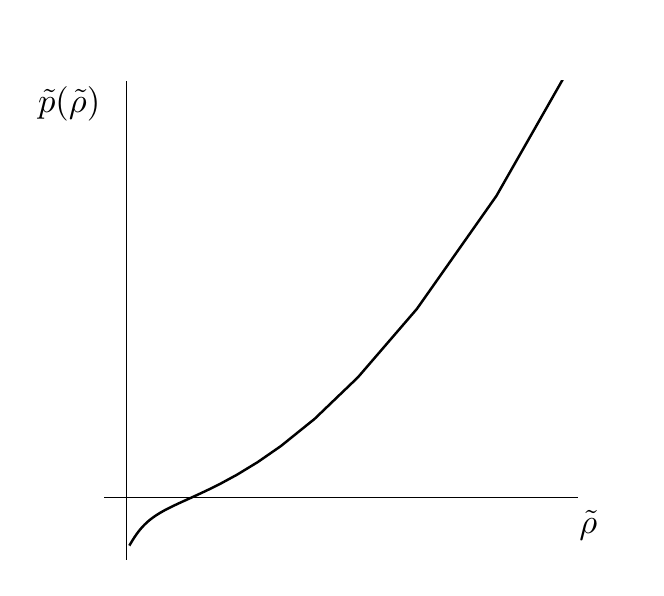}   }
\hfill
{   \includegraphics[height=4cm]{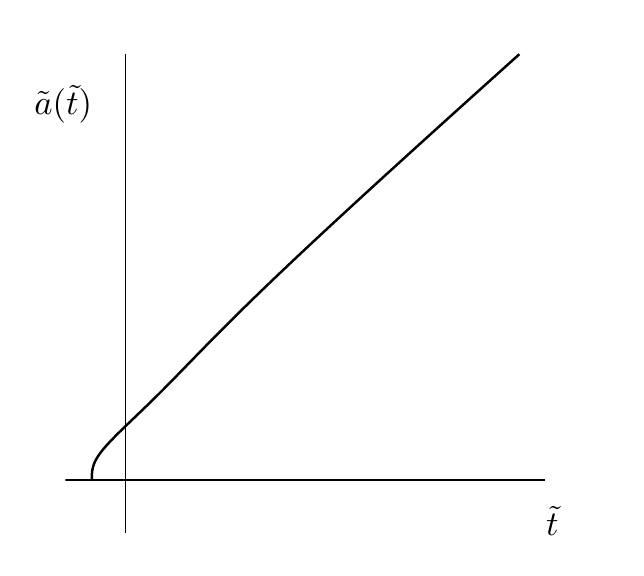}   }
   \caption{\small\it The graphs show in order:\protect\\ 
   a) the EoS $\tilde p(\tilde \rho)$ of the effective sources \protect\\ 
   b) the evolution of the scale factor $\tilde a(\tilde t)$ of the metric $\tilde g$ as a function of the coordinate time $\tilde t$ 
   }
   \label{fig:3}
\end{figure}

At that point one can graph all quantities as a function of all the others, for example the EoS for effective matter $\tilde p(\tilde \rho)$ (see Figure \ref{fig:3}.a) and 
the evolution of the scale factor of $\tilde g$ as a function of $\tilde t$, i.e.~$\tilde a(\tilde t)$ (see Figure \ref{fig:3}.b). 

As long as the old question of which frame is physical, if $g$ or $\tilde g$, the issue is simply ignored since the quantities are all physical in a sense, just some physical structures are associated to $g$, some to $\tilde g$. The issue is also meaningless since the two metrics after all are one a function of the other, so that they both equally are physical or unphysical in a sense.

\acknowledgments
We wish to thank N.Fornengo (University of Torino) and James Edholm (Lancaster University) for comments and discussions.
We also thank the (anonymous) referee for the stimulating comments about the first version of the manuscript.

This article is based upon work from COST Action (CA15117 CANTATA), supported by COST (European Cooperation in Science and Technology).
We acknowledge  the contribution of INFN (IS-QGSKY), the local research project {\it Metodi Geometrici in Fisica Matematica e Applicazioni (2017)} of Dipartimento di Matematica of University of Torino (Italy). 
This paper is also supported by INdAM-GNFM.


\vfill\eject

\end{document}